\documentstyle[aps,pre,preprint]{revtex}
\begin{document}
\draft

\title{The Fermi-Pasta-Ulam problem revisited}

\author{Lapo Casetti\cite{lapo}}
\address{Scuola Normale Superiore, Piazza dei Cavalieri 7, 56126 Pisa, Italy}

\author{Monica Cerruti-Sola\cite{monica} and Marco Pettini\cite{marco}}
\address{Osservatorio Astrofisico di Arcetri, Largo E. Fermi 5, 50125 
                                                              Firenze, Italy}
\author{E.G.D. Cohen\cite{eddie}}
\address{The Rockefeller University, 1230 York Avenue, New York, 
NY 10021-6399, USA} 

\date {\today}

\maketitle

\begin{abstract}
\medskip
The Fermi-Pasta-Ulam $\alpha$-model of harmonic oscillators with cubic
anharmonic interactions is studied from a statistical mechanical point of
view. Systems of $N= 32$ to 128 oscillators appear to be large enough to
suggest statistical mechanical behavior. A key element has been a comparison
of the maximum Lyapounov coefficient $\lambda_{max}$ of the FPU $\alpha$-model
and that of the Toda lattice. For generic initial conditions, $\lambda_{max}(t)$
is indistinguishable for the two models 
up to times that increase with decreasing energy (at fixed $N$).
Then suddenly a bifurcation appears, which can be discussed in relation
to the breakup of regular, soliton-like structures. After this bifurcation, 
the $\lambda_{max}$ of the FPU model appears to approach a constant, 
while the $\lambda_{max}$ of the Toda lattice appears to approach zero, 
consistent with its integrability.
This suggests that for generic initial conditions the FPU $\alpha$-model is
chaotic and will therefore approach equilibrium and equipartition of energy.
There is, however, a threshold energy density $\epsilon_c(N)\sim 1/N^2$,
below which trapping occurs; here the dynamics appears to be regular, 
soliton-like
and the approach to equilibrium - if any - takes longer than observable
on any available computer. Above this threshold the system appears to behave in
accordance with statistical mechanics, exhibiting an approach to equilibrium
in physically reasonable times. 
The initial conditions chosen
by Fermi, Pasta and Ulam were not generic and below threshold and would have
required possibly an infinite time to reach equilibrium.
The picture obtained on the basis of $\lambda_{max}$ suggests that neither the
KAM nor the Nekhoroshev theorems in their present form are directly relevant
for a discussion of the phenomenology of the FPU $\alpha$-model presented here.

\end{abstract}
\narrowtext

\section{Introduction}

 Few problems have been studied so extensively over the last decades as the
one devised originally by E. Fermi, J. Pasta and S. Ulam (FPU) in 1954 
\cite{FPU}. 
Their purpose was to check numerically
that a generic but very simple non-linear many particle dynamical system 
would indeed behave for large times as a statistical mechanical system, that 
is it would approach equilibrium. 
In particular their purpose
was to obtain the usual equipartition of energy over
 all the degrees of freedom of a system,
 for generic initial conditions.
To their surprise, for the system FPU considered -  a one dimensional 
anharmonic chain of 32 or 64 particles with fixed ends and in addition 
to harmonic, cubic 
($\alpha$-model) or quartic ($\beta$-model) anharmonic forces between 
nearest neighbors - this was not observed. A variety of manifestly 
non-equilibrium and non-equipartition behaviors was seen, including 
quasiperiodic recurrences to the initial state.
In fact, a behavior reminiscent of that of a dynamical system with few degrees
of freedom was found, rather than the expected statistical mechanical behavior.
The duration of their calculations varied between 10000 and 82500 computation 
steps.
 These results raised the fundamental question about the validity or at least 
the generally assumed applicability of statistical mechanics to non-linear 
systems of which the system considered by FPU seemed to be a typical example.
Most of the attempts to clarify this difficulty have approached the problem 
as one in dynamical systems. These analyses have revealed many very 
interesting properties of the FPU system and uncovered a number of possible
explanations for the resolution of the observed conflict with statistical
 mechanics. The classical explanations are: {\it i)} the survival of invariant
tori in the phase space of a quasi-integrable system (KAM theory) \cite{KAM},
{\it ii)} the existence of Zabusky and Kruskal's solitons in the KdV continuum
limit \cite{Zabusky,Zabusky1}, {\it iii)} the existence of an order-to-chaos
transition \cite{Izrailev}.
 However, we do not believe that the problem has as yet been entirely 
resolved. In particular, 
it is the purpose of this paper to try to clarify the 
problem from a statistical mechanical point of view. That is, we will try to 
exhibit the reasons why this apparently bona fide statistical mechanical 
system did not behave as such and, in particular, what in our opinion the 
significance of this apparent failure is for the general validity of 
statistical mechanics.

There are a number of obvious questions related to the unstatistical 
mechanical behavior observed by FPU, which all address the generic nature of 
their results:
\smallskip 
\begin{description}
\item{a)} Was their time of integration long enough?
 \smallskip
\item{b)} Was their dynamical system of $N=32$ or $64$ particles in one 
dimension 
large enough, i.e. possessing a sufficient number of degrees of freedom, to 
qualify as a statistical mechanical system?
\smallskip 
\item{c)} Were the recurrence phenomena (to within $3\%$) observed by FPU, 
transient or generic, i.e. possibly related to a Poincar\'e recurrence time?
\smallskip 
\end{description}
The search for answers to these questions made the work of FPU very 
seminal, spawning many new developments and connections in the theory of 
nonlinear dynamical systems, such as the connection with continuum models 
based on the Kor\-teweg - de Vries equation, leading to solitons \cite{Zabusky},
heavy  breathers  etc., or with few degrees of freedom models like 
the H\'enon-Heiles and the Toda lattice \cite{Lunsford}.
 For a recent review we refer to \cite{Ford}. 

The approach to equilibrium of the FPU - $\beta$ model was studied extensively
for various classes of initial conditions by Kantz {\it et al.}
\cite{KLR} and recently by De Luca {\it et al.} \cite{DLR}.
 A very detailed picture has emerged from their work, as to the behavior 
of the FPU-$\beta$ model in its dependence on non-equilibrium initial 
conditions as well as the role played by low frequency and  
 high frequency mode-mode couplings 
\cite{Poggi} during its time evolution.
Two threshold energies were identified, but the connection of all the 
very interesting and detailed information obtained with the generic statistical 
mechanical behavior of the FPU-$\beta$ model remains unclear so far. 

Thus, although the effort to resolve the so-called FPU problem has led to 
enormous advances in our  
understanding of non linear dynamical systems, it has not yet, in our opinion, 
led to 
a full evaluation of the statistical mechanical relevance of the FPU paradox, 
i.e. FPU's original question has not yet really been answered.

It is commonly asserted \cite{Ford} that the KAM theorem provides
 the essential answer to FPU's observations, i.e., for sufficiently small
nonlinearities and a class of initial conditions living on non-resonant tori, 
the FPU system behaves like 
 an integrable system and is represented by deformed tori in phase space. With 
increasing strength of the non-linearities, a progressive chaotic behavior 
appears, which would ultimately lead to the expected approach to equilibrium 
and equipartition \cite{nota1}.

Even though there are regular regions in phase space, the existing estimates
based on the KAM theorem are qualitatively different from what we found, 
indicating that the physics of the FPU model is quite different from what 
is contained in these estimates.
This makes us believe, on the basis of our numerical simulations, 
that it are the very special initial conditions chosen by Fermi and 
collaborators that make their system belong to a regular region of phase space.
If they
had chosen an initial excitation ten times larger they could have observed
equipartition.
This takes place via a disappearence with increasing initial amplitude
of a threshold, whose $N$-dependence is entirely different from that estimated 
in the KAM theory framework. 
Hence the source of FPU's failure is the fortuitous 
choice of initial conditions in a regular region of phase space, below this 
critical energy.

Thanks to the power of modern computers, we have considerably extended 
the calculations performed in the past by various authors. We have been 
able to reconcile different, and sometimes contradictory, aspects of the FPU
dynamics, finding that regular regions and a large "chaotic sea" can
coexist in phase space. The lack of equipartition in the original FPU 
experiment is not representative of a global property of phase space: 
apparently regular, soliton-like 
structures, similar to those of Zabusky and Kruskal, have a very long,
possibly infinite, life-time below a stochasticity threshold, whereas, above 
the same threshold, they have only a finite life-time.

By choosing more physically
generic initial conditions, i.e random positions and momenta, a threshold 
energy for the onset of chaos is detected, showing a rather 
strong tendency to vanish at increasing number of degrees of freedom.
Thus we have found strong evidence in support of the point of view that the
so-called "FPU-problem" does not invalidate the (generic) approach to
equilibrium and the validity of  
equilibrium statistical mechanics. On the other hand the existence of
long living initial states far from equilibrium,
 may well have
interesting, non trivial physical implications.

\medskip
\section{Model and results}
\medskip
We have considered a one dimensional lattice of unit mass particles
interacting via nearest-neighbor forces with unit harmonic coupling constant, 
with fixed end-points 
($q_1=q_{N+1}=0$) and described by the Hamiltonian
\begin{equation}
H({\bf p},{\bf q}) = \sum_{k=1}^N\left[ \frac{1}{2}p_k^2 + \frac{1}{2} 
(q_{k+1} - q_k)^2 + \frac{\alpha}{3}(q_{k+1} - q_k)^3\right]
\label{HFPU}
\end{equation}
which is known as the FPU $\alpha$-model.
We can think of (\ref{HFPU}) as the first anharmonic approximation to
physical interatomic potentials of the Morse or Lennard-Jones type.

The cubic term is obviously responsible for the energy exchange among the 
normal modes of the harmonic part of (\ref{HFPU}).
The normal mode coordinates, obtained by a standard orthogonal 
transformation, read 
\begin{equation}
Q_k=\left({\frac{2}{N}}\right)^{1/2} \sum_{i=1}^N q_i \sin \frac{i k \pi}{N}
\label{norm-modes}
\end{equation}
and diagonalize the harmonic part of (\ref{HFPU}).

The Hamiltonian in these new coordinates becomes
\begin{equation}
H({\bf P},{\bf Q}) = \sum_{k=1}^N \left[ \frac{1}{2}P_k^2 + 
\frac{1}{2}\omega_k^2 Q_k^2 + \alpha \sum_{k^\prime,k^{\prime\prime}=1}^N
C(k,k^\prime,k^{\prime\prime})Q_kQ_{k^\prime}Q_{k^{\prime\prime}}\right]
\label{hamilt-nm}
\end{equation}
where $\omega_k=2\sin \left(\frac{k\pi}{2N}\right)$ is the frequency of the 
$k$-th normal mode and $P_k=\dot Q_k$ are the 
 conjugated momenta. The natural unit of time with the choice of the units of 
Eq. (\ref{HFPU}), is given by the inverse of the 
fastest frequency of the harmonic part of (\ref{HFPU}): $T_{min}={2\pi}/
{\omega_{max}}\equiv \pi$. In what follows $t=1$ corresponds to $1/\pi$ of 
this fastest linear period.

We have numerically integrated this model by means of a very efficient 
third order 
bilateral symplectic algorithm \cite{Casetti} that ensures a faithful
representation of a Hamiltonian flow and, with the adopted values of the 
rather large time
step ranging from $0.01$ to $0.1$, 
keeps the total energy $E$ constant within an average fluctuation level of
$\Delta E/E\simeq 10^{-8}$ without drift.  
Such a high precision in numerical integration makes the outcome of
the very long runs that are reported in the following reliable.
The coupling constant was $\alpha =0.25$. In order to give an idea of the
computational effort that was necessary to obtain the results reported in
this paper, let us mention that  the computation of the largest Lyapunov
exponents $\lambda_1$ at low energy typically required integration times in the 
interval $10^7$ -- $10^9$ natural units of time: the 
longest runs lasted about $60$ hours
of CPU time on an $HP9000/735$ computer (about the same CPU time would be
necessary on a CRAY Y-MP computer). The CPU time 
amounted to about $4000$ hours on the following computers: 
$HP 9000/735$,
$HP 9000/715$, $Sun$ $SPARC10$, $Sun$ $SPARC5$, $Digital$
 $Alphaserver$ $2000$ $4/200$.

For what concerns the initial conditions, we begin by choosing single mode
excitations as Fermi and collaborators did in their original experiment, i.e.
the initial displacements of the particles from their equilibrium positions
are given by the fundamental mode:
\begin{equation}
q_i(0)= A \sin \left( \frac{2\pi n i}{N}\right)~,~~~i=1,\dots ,N~~,
\label{c.i.}
\end{equation}
and the initial momenta $p_i(0)=0$ for $i=1,\dots ,N$.
Fermi {\it et al.} used $N=32$, $A=1$, and mode number $n=1$.

 Now the main question  is: if 
we repeat the original experiment, do we have any chance to find equipartition
or any clear tendency to it by using present day fast computers?

To answer this question we cannot blindly make the longest integration of
the trajectories of the Hamiltonian (\ref{HFPU}), that we
can afford on a very fast computer: the absence of equipartition, 
 even after a very long integration time, would not be by itself conclusive
for the non existence of equipartition for this model. Rather, one should make
an estimate 
of the equipartition time at $A=1$ and $N=32$, and determine
the $A$-dependence of this time for larger values of $A$: a large
initial excitation amplitude makes the anharmonicity of the system larger,
hence the mode-mode couplings stronger, so that  
a faster relaxation to equipartition can be expected.  
\smallskip
\subsection{Detecting energy equipartition}
\smallskip
A possible method to numerically detect equipartition of energy makes use of the
spectral entropy \cite{LPRSV1} defined by
\begin{equation}
{\cal S}(t) = - \sum_{k=1}^{N/2} w_k(t)\,\ln w_k(t)
\label{entrspett}
\end{equation}
where the weights $w_k$ are given by the fraction of the total harmonic energy
$E_k=\frac{1}{2}(P_k^2+\omega_k^2Q_k^2)$ in the $k$-th normal mode \cite{nota2}
\begin{equation}
w_k(t) = \frac{E_k(t)}{\sum_i E_i(t)}
\label{pesi}
\end{equation}
so that ${\cal S}(t)$ reaches its maximum value when equipartition is attained.
This entropy can be normalized as follows
\begin{equation}
\eta (t) = \frac{{\cal S}(t) - {\cal S}_{max}}{ {\cal S}(0)- {\cal S}_{max}}~,
\label{eta}
\end{equation}
hence $\eta =1$ detects a ``freezing'' of the initial condition and $\eta =0$
detects equipartition. 

By following the time relaxation of $\eta (t)$ we have obtained some 
estimates of the equipartition time $\tau_{{}_E}$ at 
values of the initial excitation amplitude $A$ ranging from $3$ to $11$. 
Note that $A$ cannot be too large, since then the cubic
part of the potential becomes repulsive and the phase space trajectories
are ``runaway'' trajectories, nor can $A$ be too small, since then the 
relaxation times
become too large to be determined by our computations.
However, two major difficulties arise with the $\eta$-method:  
{\it i)}  contrary to 
previously investigated systems \cite{PettiniLandolfi,PettiniCerruti},
the relaxation pattern of $\eta (t)$ does not show clearly when equipartition
sets in, so that  the equipartition time is a fuzzy quantity, 
{\it ii)} if we somehow {\it define} the equipartition time by, for example,
 measuring
the time needed for $\eta (t)$ to drop below a threshold value, say 
$\eta =0.1$,
we find that  $\tau_{{}_E}(\epsilon )\sim \epsilon^{-3}$, ($\epsilon =E/N$ is
the energy density).   
For the FPU case ($A=1$, $\epsilon =0.00241$), we will show (see Section 
\ref{trapping}) that the equipartition time 
is far too long a time for numerical tests, since it would amount to 
about a year's CPU time.
\smallskip
\subsection{Phase space trapping}
\label{trapping}
\smallskip
We have been able to overcome these difficulties by focussing on the
development of chaoticity in the time evolution of the system rather than
on the attainment of  equipartition. 

The natural way of characterizing chaotic motions is to compute the largest
Lyapunov exponent $\lambda_1$. Let us briefly remember its definition. Let
\begin{equation}
\dot x^i = X^i(x^1\dots x^N)~,~~~~~i=1,\dots ,N
\label{eq1}
\end{equation}
be a given dynamical system, and denote by
\begin{equation}
\dot\xi^i = \sum_{k=1}^N {\cal J}^i_k[ x(t)]\, \xi^k~,~~~~~i=1,\dots ,N 
\label{eq2}
\end{equation}
its tangent dynamics equation, 
$[{\cal J}^i_k]=[\partial X^i/\partial x^k]$ is the Jacobian matrix of $[X^i]$,
 then the largest Lyapunov
exponent $\lambda_1$ is defined by
\begin{equation}
\lambda_1 = {\displaystyle\lim_{t\rightarrow\infty}}~\frac{1}{t}\ln \frac
{\Vert\xi (t)\Vert}{\Vert\xi (0)\Vert}~.
\label{eq3}
\end{equation}
By setting $\Lambda [x(t),\xi (t)]\equiv \xi^T\,{\cal J}[x(t)]\, 
\xi /\,\xi^T\xi =\xi^T {\dot\xi} /\xi^T \xi 
=\frac{1}{2}\frac{d}{dt}\ln (\xi^T\xi )$, 
$\lambda_1$ can be formally expressed as a time average
\begin{equation}
\lambda_1 ={\displaystyle\lim_{t\rightarrow\infty}}~\frac{1}{2t}\int_0^t 
\,d\tau \,\Lambda [x(\tau ), \xi (\tau )]~~~
\label{eq4}
\end{equation}
which, in practice, is evaluated by computing \cite{BGS}
\begin{equation}
\lambda_1 (t_{\cal N}) = 
\frac{1}{{\cal N}\Delta t} \sum_{n=1}^{\cal N} \ln \left( 
\frac{\|\xi(t_n )\|}{\|\xi (t_{n-1})\|}\right) \, ,
\label{LEav}
\end{equation}
where $t_n = n\Delta t$ ($\Delta t$ is some time interval), up to a final time 
$t_{\cal N}$, $n={\cal N}$, 
such that $\lambda_1 (t_{\cal N})$ has converged to a reasonably
asymptotic value \cite{notaly}. A positive asymptotic value of $\lambda_1$
obviously detects a chaotic dynamics, whereas $\lambda_1(t)\sim t^{-1}$
corresponds to a non-chaotic dynamics.

We have exploited the existence  
of an integrable model, the Toda lattice, 
close to the FPU-$\alpha$ model. The Toda lattice Hamiltonian reads, in the 
same units as used in Eq.(\ref{HFPU}): 
\begin{equation}
H({\bf p},{\bf q}) = \sum_{k=1}^N \left[\frac{1}{2}p_k^2 + \frac{1}{4\alpha^2}
\left\{\exp [-2\alpha (q_{k+1} - q_k)] + 2\alpha (q_{k+1} - q_k) -1\right\}
\right]~;
\label{HToda}
\end{equation}
and the power series expansion of its potential coincides, up to third order,  
with the potential part of (\ref{HFPU}).

As the Toda lattice Hamiltonian describes an integrable system, the numerical 
computation of the time behavior of the largest Lyapunov exponent 
$\lambda_1(t)$ must reveal the non-chaotic property of its dynamics.
In particular, one might wonder 
whether some information can be obtained by comparing
$\lambda_1^{{}^{FPU}}(t)$ with $\lambda_1^{{}^{Toda}}(t)$ for the same initial 
conditions, i.e. $\{ q_i(0), p_i(0)\}$, $i=1,\dots ,N$.
This turns indeed out to be the case.
In Fig. 1 we report, as an example, 
$\lambda_1^{{}^{FPU}}(t)$ and $\lambda_1^{{}^{Toda}}(t)$ in the case $N=32$ and
$\epsilon =0.0217$ for an initial condition with $A=3$ in Eq. (\ref{c.i.}). Up  
until $t\simeq 2\cdot 10^{5}$ these two functions are so close to each 
other that the two are virtually indistinguishable.
Then - suddenly - they separate at $t\geq 4\cdot 10^{5}$:
 $\lambda_1^{{}^{Toda}}(t)$
continues its decay toward zero, whereas $\lambda_1^{{}^{FPU}}(t)$ tends
to converge to a non-vanishing value. This makes possible to define clearly
what a trapping time in a regular region of phase space is; moreover, its
numerical determination is unambiguous, as it can be deduced by simply
looking at Fig. 1. 
We attribute this dramatic difference to the un-trapping of the FPU system
from its regular region in phase space by escaping to the chaotic component
of its phase space.

The peculiar behavior of $\lambda_1^{{}^{FPU}}(t)$ suggests therefore
 that non-integrable
motions, originated by one-mode initial excitations, after a transient, possibly long, trapping
in a regular region of phase space, enter its chaotic component. The chaotic
component of phase space, by the
Poincar\'e-Fermi theorem \cite{pf}, is connected, so that we may well expect 
that equipartition is eventually attained
on a finite, albeit possibly very long, time scale.

The precise nature of the trapping mechanism is unclear to us. 
One could either think that the trajectories stick close to a KAM torus 
during
the trapping time, or that - for example - they are geodesics  of a bumpy 
manifold to which 
an $N$-torus has been differentiably glued by a tiny "bridge"; in that 
case a geodesic - originating at any point of the $N$-torus - is locally stable
until it finds a path to escape from the torus.
Another possible mechanism of trapping and escape is discussed in the 
Discussion Section under point $5$, in the light of the data discussed 
in the sequel of the present Section.

In conclusion, once we observe that a trajectory enters the chaotic 
component of 
phase space, we are allowed to think that equipartition will be eventually 
attained.  
In Fig. 2 the relaxation times to equipartition $\tau_E(\epsilon )$
and the trapping times $\tau_T(\epsilon )$, 
evaluated for the FPU-$\alpha$ model, are shown.
The results refer to $N=32$ and initial conditions given by Eq.(\ref{c.i.}).
The initial excitation amplitudes $A$ range from $3$ to $11$.
The equipartition times  appear to be between one and two orders of magnitude
larger than the trapping times. If we extrapolate \cite{nota3} the
equipartition time to the case of FPU, we find $\tau_{{}_E}\simeq 4\cdot 10^ 
{10}$ (as is indicated in Fig. 2 by an asterisk).

Since we can consider the FPU $\alpha$-model as a third
order expansion of a Lennard-Jones potential around its minimum, 
we can tentatively extrapolate the equipartition times reported in Fig. 2
to a macroscopic system in three dimensions.
As an example, 
we can roughly estimate what the physical equipartition time could be
for a {\it classical} Xenon crystal at zero temperature and only 
one normal mode initially excited.
At $\epsilon =0.01$, $\tau_{{}_E}\sim 10^{9}$ proper times (see Fig. 2)
corresponds to about $10^{-3} s$, using as proper time, 
obtained with standard Lennard-Jones parameters for Xenon, $2.4\cdot 10^{-12} 
s$.
\smallskip
\subsection{Stochasticity threshold}
\smallskip
We have evaluated 
the trapping times $\tau_{{}_T}(\epsilon ,N)$  for the FPU-$\alpha$ model    
at different values of both the energy density $\epsilon$
and of the number of degrees of freedom $N$. 
When $N$ was varied, we kept the wavelength of the initial excitation constant
(i.e. $n$ in Eq.(\ref{c.i.}) is taken proportional to $N$: $n=1$ at $N=32$, 
$n=2$ at $N=64$ and $n=4$ at $N=128$),
and we excited only one mode at $t=0$. The results are reported in Fig. 3.
With decreasing $\epsilon$, first $\tau_{{}_T}$ tends to increase 
monotonically, then, abruptly, it displays an apparently divergent behavior.
In fact, when reaching critical threshold values, a very small change in
$\epsilon$ suddenly gives an extremely steep increase in $\tau_{{}_T}$.
This very steep increase of $\tau_{{}_T}$ with decreasing $\epsilon$ suggests
at least a very narrow bottleneck in phase space, through which the system 
can only escape
with great difficulty. We assume that this bottleneck is not an
insurmountable barrier, and for that reason, as well as to conform with      
previous use, we will call it a {\it threshold}.
This is consistent with the results of Fig. 4, where the largest Lyapunov 
exponents  are reported for the same cases as in Fig. 3. 
We have used a cut-off of the integration time at $t=4.3\cdot 10^8$.
After such a long time - when $\epsilon$ was smaller than the threshold 
value - no separation between $\lambda^{Toda}_1(t)$ and $\lambda^{FPU}_1(t)$
has been observed: the values of $\lambda^{FPU}_1$ at $t=4.3\cdot 10^8$ 
(endpoints
of broken lines in Fig. 4, marked by arrows) are taken as upper bounds
 for the FPU-Lyapunov exponents and
$t=4.3\cdot 10^8$ is taken as a lower bound for the trapping time of the 
FPU-phase point (endpoints of the broken lines in Fig. 3, marked by arrows).
Both $\lambda_1(\epsilon ,N)$ (from here on, by $\lambda_1$ we mean 
$\lambda^{FPU}_1$) and
$\tau_{{}_T}(\epsilon ,N)$ strongly suggest the existence of an 
$N$-dependent threshold value of the energy
(density), above which the motion is chaotic and below which the trajectories
appear to be trapped in a regular region of phase space [{\it 
Stochasticity Threshold} (ST)]. 

The following approximate relationship between $A$ and $\epsilon$ holds:
$\epsilon\simeq 0.00241 A^2$ therefore the threshold amplitudes $A_c$ leading 
to chaos are: $A_c(N=32)\simeq 1.62$,  $A_c(N=64)\simeq 1.42  $, 
$A_c(N=128)\simeq 1.28$, respectively.

The dotted line at $\epsilon = 0.00241$ corresponds
to the initial excitation amplitude $A=1$ of FPU's original paper. 
We see that Fermi and coworkers chose an
initial condition well below this ST \cite{nota3}. Figure 3 shows 
 that if they had taken 
ten times larger amplitude, they would have observed equipartition 
during the integration time they used.
This appears to us to be the explanation of the lack of statistical mechanical 
behavior observed in the original FPU numerical experiment.

A few more comments on these results: {\it i)} Fig. 3 suggests that with 
increasing $N$, $\tau_{{}_T}(\epsilon ,N)$ probably becomes less dependent 
on $N$ [in fact the values of $\tau_{{}_T}(\epsilon ,N=64)$ are closer to those
of $\tau_{{}_T}(\epsilon ,N=128)$ than to those of $\tau_{{}_T}(\epsilon 
,N=32)$]; {\it ii)} there is a narrow energy density interval where 
$\tau_{{}_T}$ and $\lambda_1$ oscillate, 
suggesting the transition in phase space from chaotic behavior at large 
$\epsilon$ to regular behavior at small $\epsilon$;
{\it iii)} the ST shows a weak but clear dependence on $N$  
(provided 
the wavelength of the initial excitation is kept constant). This is not
surprising because if we combine two identical systems - each with the same 
initial excitation - the composite system will have new low frequency modes
that are absent in the separate subsystems, which can be expected to 
facilitate the mechanism of energy exchange among the normal modes.

As far as we are aware, the {\it direct} evidence given in Fig. 4 
of  the existence of a ST 
at $N\gg 2$, obtained 
through the behavior of
the largest Lyapounov exponent, 
was never found before in nonlinear Hamiltonian systems. 
An {\it indirect} suggestion of its existence 
(through a ``freezing'' of the decay of the spectral
entropy)
has been given in \cite{KLR} for the FPU $\beta$-model.
\smallskip
\subsection{Coexistence of order and chaos}
\smallskip
A question now arises: does such a threshold refer to a global property of
the constant energy surface $\Sigma_{{}_E}$ or is it, rather, a local property
of $\Sigma_{{}_E}$ sensitive to the initial condition?

In order to answer this question we have considered also the following
initial conditions at $N=32$: {\it i)} $A\neq 0$ [the fundamental mode - Eq.
(\ref{c.i.}) - is excited with amplitudes ranging from $A=0.7$ to $A=5.5$ ] 
and $p_i(0)\neq 0$ ($i=1,\dots ,N$) are gaussian
random numbers with zero mean and standard deviation (``temperature'')
equal to $0.001$;
{\it ii)} $A=0$, $q_i(0)=0$ ($i=1,\dots,N$) and $p_i(0)$ ($i=1,\dots ,N$) 
are randomly
chosen according to a gaussian distribution. 

In the first case, a fraction
of the energy in the initial excitation is given to all the normal modes
of the system: in so doing, we are displacing the starting point on 
$\Sigma_{{}_E}$, proportional to the magnitude of the
standard deviation of the noisy component of the initial excitation. This is
intended to provide some information about the extension of the regular
 region(s) in phase space.
If all, or almost all, the energy of the initial excitation 
is concentrated in one or a few modes, then we are dealing with a 
non-equilibrium
initial condition; the existence of a ST entails then that a system, 
prepared in
a non-equilibrium state below such a threshold, will appear never to 
attain equipartition
of the energy of the initial excitation. 
The noisy component has a self evident physical meaning
related to the impossibility of preparing any physical system in a perfectly
ordered initial state: at non-zero temperature some randomness in the initial
conditions is unavoidable.

The second choice - completely random initial conditions - mimicks 
a physical situation that corresponds to a crystal prepared at an assigned
temperature (i.e. mean kinetic energy per degree of freedom) at thermal
equilibrium. 

In Fig. 5 the effect of changing the initial conditions according to the above
prescriptions is shown. Here $N=32$, the squares refer to $A\neq 0$ and
no random component, the  asterisks refer to only random initial excitations
and the star-like squares refer to $A\neq 0$ plus a random component.

In each case a threshold energy (or equivalently energy density, since $N$ is 
fixed) is found. 
At $\epsilon > 0.01$, the uncertainties in the determination of $\lambda_1$ 
are of the order of the 
size of the symbols used; at $\epsilon < 0.01$, an estimate is difficult
because of unpredictable fluctuations of $\lambda_1(t)$ that could be reduced
only by prohibitively long integration times. Nevertheless,   
the information given
by Fig. 5 is unambiguous. Down to $\epsilon \simeq 10^{-2}$ all the values of
$\lambda_1(\epsilon )$, obtained with different initial conditions, crowd along
the same line; below $\epsilon\simeq 10^{-2}$ 
the values of  $\lambda_1$, obtained with different initial conditions, 
separate and exibit a coexistence of regular and chaotic regions on the
constant energy surface in phase space, whose details depend on the 
initial conditions.
Below $\epsilon\simeq 1.8\cdot 10^{-3}$, even with only random initial
conditions, the motions are regular (star-like squares).

This tells us that the phase space undergoes some important structural
change as a function of the energy, in analogy with what is observed in 
two-degrees-of-freedom systems,
like the H\'enon-Heiles model \cite{Henon}, where 
fully developed chaos, a coexistence of regular and 
chaotic regions of phase space, or    
only regular trajectories  are observed, depending on the value of the energy. 

\smallskip
\subsection{N-dependence of the stochasticity threshold}
\smallskip
An important question is the stability or instability of 
the stochasticity threshold with $N$. To this end we have numerically 
determined  
$\lambda_1(\epsilon ,N)$ at $N=8, 16, 32, 64$ always choosing random initial
conditions, i.e. $\{q_i(0)=0, p_i(0)= r_i\}$, $i=1,\dots ,N$, with $r_i$ 
gauss-distributed
random numbers with zero mean and variance $\sqrt{2\epsilon}$ (after the
assignement of the random values $r_i$, the momenta are adjusted, by rescaling 
them, in order to obtain a fixed initial $\epsilon$).

In Fig. 6 the outcome of these computations is reported.
The endpoints of the broken lines have the same meaning as already discussed 
above. At large $\epsilon$ there is a tendency of all the sets of points to
join. This fact is most evident for $N=32$ and $N=64$ which have a line
segment in common and then separate at small $\epsilon$: the larger $N$, 
the smaller the $\epsilon$ at which the separation occurs. 

At each $N$, we take as rough estimate of the stochasticity threshold 
$\epsilon_c$, the value of $\epsilon$ at the midpoint between the two lowest
points on each curve, because the lowest point (marked by an arrow)
 is presumably below threshold.

Figure 7 then shows that for these threshold values $\epsilon_c$ plotted 
{\it vs.} $N$ holds: 
$\epsilon_c(N)\sim 1/N^2$. 
This result is interesting for the following reasons:
\begin{description}
\item{ i)} the
threshold values vanish sufficiently fast with increasing $N$, 
that the existence of regular regions of phase space below $\epsilon_c$ does
not constitute a problem for {\it equilibrium} statistical mechanics. 
In fact, it appears that $N=32,\dots ,128$ is not too small to obtain 
indications, if not ``confirmation'', of the statistical
mechanical behavior of the FPU-$\alpha$ system.
\smallskip
\item{ ii)} both the values of $\epsilon_c$ 
and the $N$-dependence $\epsilon_c(N)\sim 1/N^2$ can hardly be 
explained  on the basis of the best available estimate of the 
perturbation amplitudes for which a positive
measure of the regular KAM regions in phase space exists:  
$\mu < \mu_c\sim a\exp (-bN \ln N)$ \cite{stimekam}, where
$\mu$ measures the relative strength of the anharmonic 
to the  harmonic part of a given Hamiltonian 
($\mu$ depends on $\epsilon$), $\mu_c$ is a threshold value, $a$ and $b$
are constants. There are also power-law estimates for $\mu_c(N)$, i.e.
$\mu_c(N)\sim N^{-\delta}$, obtained in the context of KAM theory \cite{wayne},
however - at present - $\delta$ is still very large: $\delta\simeq 160$.
\smallskip 
\item{iii)} the bounds on the threshold value 
$\frac{1}{4\alpha^2N}\leq\epsilon_c
\leq\frac{1}{2\alpha^2N}$ found by Enz {\it et al.} for the FPU-$\alpha$ model 
\cite{Enz} predict an  
$N$-dependence which is in much 
better agreement with our results than the exponential drop with $N$ mentioned
in the previous point.
\smallskip
\end{description}
\medskip
\section{Discussion}
\label{discussion}
\medskip
We conclude with the following remarks.

1. It is worth mentioning that recently the existence of an equipartition 
threshold vanishing at increasing $N$ has also been reported for the 
FPU-$\beta$
model \cite{KLR,Kantz,Lichtenberg}. It is inappropriate to compare these 
results 
{\it quantitatively} with ours: our model is different and the evidence for a 
stochasticity threshold for the FPU $\beta$-model 
has been obtained indirectly, through the opening of
a local trap in phase space, which prevented equipartition. However, we can say
that both results are in {\it qualitative} agreement: threshold effects, 
either concerning transient dynamics to equipartition
of non-equilibrium initial conditions (studied through spectral entropy) 
or concerning the dynamics of equilibrium initial conditions 
[studied through $\lambda_1(\epsilon ,N)$], vanish at increasing $N$. 

A qualitative agreement about the vanishing with $N$ of the critical energy
to get chaos is reported in a recent paper on the FPU $\alpha$-model
\cite{shepelyansky}.

The question of how to explain the existence and the  $1/N^2$
dependence of the stochasticity threshold reported here, remains open.

2. In this paper we report the existence of two interesting phenomena among
others:
the apparent existence of regular regions in the phase space of a 
non-integrable Hamiltonian system as is the FPU-$\alpha$ model and the 
existence of almost regular regions of phase space where the trajectories are 
trapped during long but finite times. These phenomena are reminiscent of the 
KAM and Nekhoroshev \cite{Nekhoroshev} theorems respectively. We have already
discussed throughout the paper why our results disagree with the present-day 
quantitative predictions of the KAM theorem.
Similarly, the Nekhoroshev theorem does not seem to be able to provide
an explanation of the observed phenomenology as well. In fact,  
without entering into the details of how 
a thorough comparison with our results could be made, Nekhoroshev's 
estimate of the 
lower bound of the ``trapping'' time $T$ of a trajectory close to its initial 
condition is $T\simeq \exp (c/\mu )^{\gamma (N)}$, where $c$ is a constant,
$\mu$ has the same meaning as above, and $\gamma (N)\sim 1/8N$ is the optimal
$N$-dependence \cite{Galgani} of the exponent. Hence 
for $N=32, 64, 128$, as is the case of the results of Fig. 3 for
$\tau_{{}_T}(\epsilon ,N)$, the ``Nekhoroshev trapping'' time $T$ 
is ${\cal O}(1)$ independently of $\mu$ (thus of $\epsilon$). 
Therefore, it appears that the physical mechanisms responsible for
finite time trapping in phase space, that are behind the Nekhoroshev theorem
and our numeric phenomenology, respectively, are different.     
Thus we emphasize the difference between, on the one hand, the approach to 
infinite time trapping
(KAM theorem) and to finite time trapping (Nekhoroshev
theorem) based on the description of the persistence of certain local properties
of regular regions of phase space,
and, on the other hand, our chaos-related approach, based on the comparison
of $\lambda_1(t)$ for the FPU and Toda lattices (Fig. 1). 
The behavior of $\lambda_1(t)$ suggests that
the sudden escape from the regular region occurs as if the trajectory would
eventually find a ``hole'' in its boundary. 

3. It is not out of place to note that while the ST drops to zero as 
$N\rightarrow\infty$ in the FPU $\alpha$-model, there is another 
transition phenomenon that does not 
suffer this $N$-dependence: it is the Strong Stochasticity Threshold (SST)
that concerns a transition from weak to strong chaos 
\cite{PettiniLandolfi,PettiniCerruti} which has been discovered in the FPU 
$\beta$-model and in the lattice $\varphi^4$ model. Unfortunately, this SST 
cannot be
investigated in the FPU-$\alpha$ model because the cubic potential prevents
working at too large energy.
We note, however, that the stability with $N$ of the SST makes the SST 
not only a dynamical phenomenon related to dynamical chaos, but also  
of potential interest to statistical mechanical phenomena.

4. The existence of special initial conditions that create non-trivial
dynamical behavior, though irrelevant for {\it equilibrium} statistical 
mechanics 
because of their negligible measure, could be physically relevant for transient
{\it non-equilibrium} statistical mechanics if we can conceive an operational
method to prepare a real system in such a special initial state (see for
example \cite{casettilivimacchipettini}).

5. Let us comment about the classical explanation of the non-statistical 
mechanical behavior 
of the FPU model proposed by Zabusky {\it et al.} 
\cite{Zabusky,Zabusky1}, based on the existence of soliton solutions of the 
Korteweg-de Vries (KdV) equation, derived as a special continuum 
limit of the FPU $\alpha$-model.
Their experiments were carried out over much shorter time scales than those 
that are possible nowadays.
As we have seen above, thanks to very long numerical
integrations, it turns out that, below a threshold, regular regions of phase 
space can coexist with chaotic ones.
In the light of our results, we can thus assert  that the KdV soliton solutions 
belong to the regular region of phase space of the FPU system
from which - after sufficiently long time - escape will occur to the 
chaotic component of phase space.

In Fig. 8 the patterns of the displacements $q_i$ as a function of position $i$
are shown at different times in the case $N=32$, $A=3$. For times $t$ 
below the trapping time ${\tau}_T\simeq 4\cdot 10^5$, they are 
looking as regular 
structures, apparently composed of a superposition of a small number of 
waves, which display a trapping and a soliton-like recurrence (see Fig. 8b),
similar to that observed by Zabusky and Kruskal \cite{Zabusky} and
Tuck and Menzel \cite{Tuck}, although we 
have fixed boundary conditions, as in FPU's original paper, rather than 
periodic boundary conditions as considered by Zabusky and Kruskal. 
For $t > \tau_T$ we observed a gradual untrapping or decay of these
regular features due to more and more complicated structures consisting of 
a superposition of an increasing number of different waves (radiated
by the decaying solitons \cite{Zabusky1}). Finally, 
for $t > 10^8$, an almost noisy pattern is attained (see Fig. 8c) as well as
energy equipartition (detected through the spectral entropy). 
This demonstrates clearly the compatibility of the existence
of trapping, i.e. very long lived regular solutions, and the attainment of
equipartition, albeit after possibly very long times. It is still an open 
question whether, at fixed $N$, below the Stochasticity Threshold, i.e.
at energy density such that the largest Lyapunov exponent seems to vanish
for any initial condition, the lifetime of regular solutions might actually
diverge.

However, we have found that this threshold energy density drops 
to zero as $\sim 1/N^2$. Therefore it appears to us that at sufficiently
large $N$ the existence of KdV solitons does not hinder a good statistical
mechanical behavior of the FPU system.

\acknowledgments
This work was started during a Workshop of the INFM - FORUM 
held in Florence during 
September 1995. E.G.D. C. wishes to thank the INFM - FORUM as well as the
Department of Energy under grant DE-FG02-88-ER13847 for financial support. 
We thank N.J. Zabusky for drawing our attention to some of his early work
on solitons and for some very helpful suggestions.
Thanks are due also to A.J. Lichtenberg and S. Ruffo for stimulating 
discussions. 
All the authors wish to thank the I.S.I.
Foundation, where this work has been concluded during the July 1996 Workshop on
``Complexity and Chaos'', financed by EU HC\& M Network ERBCHRX-CT940546.

The I.S.I. Foundation is also warmly acknowledged for having allowed us 
access to its computers.

\begin{figure}

\caption{ 
$\lambda_1^{{}^{FPU}}(t)$ (solid triangles) and $\lambda_1^{{}^{Toda}}(t)$
 (squares) are plotted {\it vs.} time for $N=32$ and energy density $\epsilon=
0.0217$ (which corresponds to an initial excitation amplitude $A=3$ in   
 Eq. (\ref{c.i.})). Dividing $t$ on the horizontal axis by $\pi$ gives the
time in units of the fastest period of the harmonic part of the chain, $T_
{min}=\pi$. }
\end{figure}

\begin{figure}
\caption{
 The relaxation times to equipartition (full squares) and the trapping times
 (open squares) are plotted (for the FPU-$\alpha$ model) {\it vs.} 
 the energy density $\epsilon$ for $N=32$ and the initial conditions of
Eq. (\ref{c.i.}). The initial excitation amplitudes considered are, 
from left to right, $A= 2, 3, 4, 5, 8, 10, 11$, respectively. 
The asterisk represents the extrapolation
of the equipartition time to the case $A=1$ (FPU's original paper
\protect\cite{nota3}).  }
\end{figure}

\begin{figure}
\caption{
 The trapping times $\tau_T(\epsilon,N)$ at different values of energy density
$\epsilon$
(i.e. at different values of the initial excitation amplitudes $A$), are
reported. Open squares refer to the case $N=32$ ($A$ ranges from 1.6 to 11),
solid triangles refer to $N=64$ ($A$ ranges from 1.4 to 10), open circles 
refer to $N=128$ ($A$ ranges from 1.25 to 9), respectively. 
The endpoints of the broken lines are lower bounds
for the trapping time (the cut-off of the integration time is at $t=4.3\cdot
10^{8}$). The dotted vertical line at $\epsilon=0.00241$ corresponds to the
initial excitation amplitude $A=1$ of the FPU's original paper.}
\end{figure}

\begin{figure}
\caption{
 The largest Lyapounov exponents $\lambda_1(\epsilon,N)$ are shown for
 different values of the energy density $\epsilon$ and a sine wave
initially excited [Eq.(4)].
 Open squares refer to $N=32$ and $n=1$, full triangles to $N=64$ and $n=2$, 
open circles to $N=128$ and $n=4$, respectively. 
The endpoints of the broken lines, marked by arrows,
 are upper bounds for the FPU-
Lyapounov exponents at $t=4.3\cdot10^8$ (cut-off of the integration time).
The dotted vertical line at $\epsilon=0.00241$ corresponds to $A=1$
(FPU's original paper).  }
\end{figure}

\begin{figure}
\caption{
 The largest Lyapounov exponents $\lambda_1(\epsilon)$ are plotted, for
 $N=32$, at
 different values of $\epsilon$ and different initial conditions.
 Squares refer to the cases  $A\neq 0$ and no random initial conditions,
 asterisks refer to the cases $A=0$ and random initial conditions,
 star-like squares to $A\neq0$ and random initial conditions with zero 
mean and standard deviation 0.001.
The endpoints of the broken lines, marked by arrows,
 are upper bounds for the FPU-
Lyapounov exponents at $t=4.3\cdot10^8$ (cut-off of the integration time). }
\end{figure}

\begin{figure}
\caption{
 The largest Lyapounov exponents $\lambda_1(\epsilon,N)$  are plotted
{\it vs.} the energy density $\epsilon$, for different values of $N$. 
Random initial conditions are chosen.
 Star-like polygons refer to $N=8$, crosses to $N=16$, asterisks to $N=32$,
 star-like squares to $N=64$, respectively.
The endpoints of the broken lines, marked by arrows,
 are upper bounds for the FPU-
Lyapounov exponents at $t=4.3\cdot10^8$ (cut-off of the integration time). }
\end{figure}

\begin{figure}
\caption{
 The values of the stochasticity thresholds $\epsilon_c$ are plotted 
for different values of $N$ (for the estimate of $\epsilon_c$, see text). }
\end{figure}

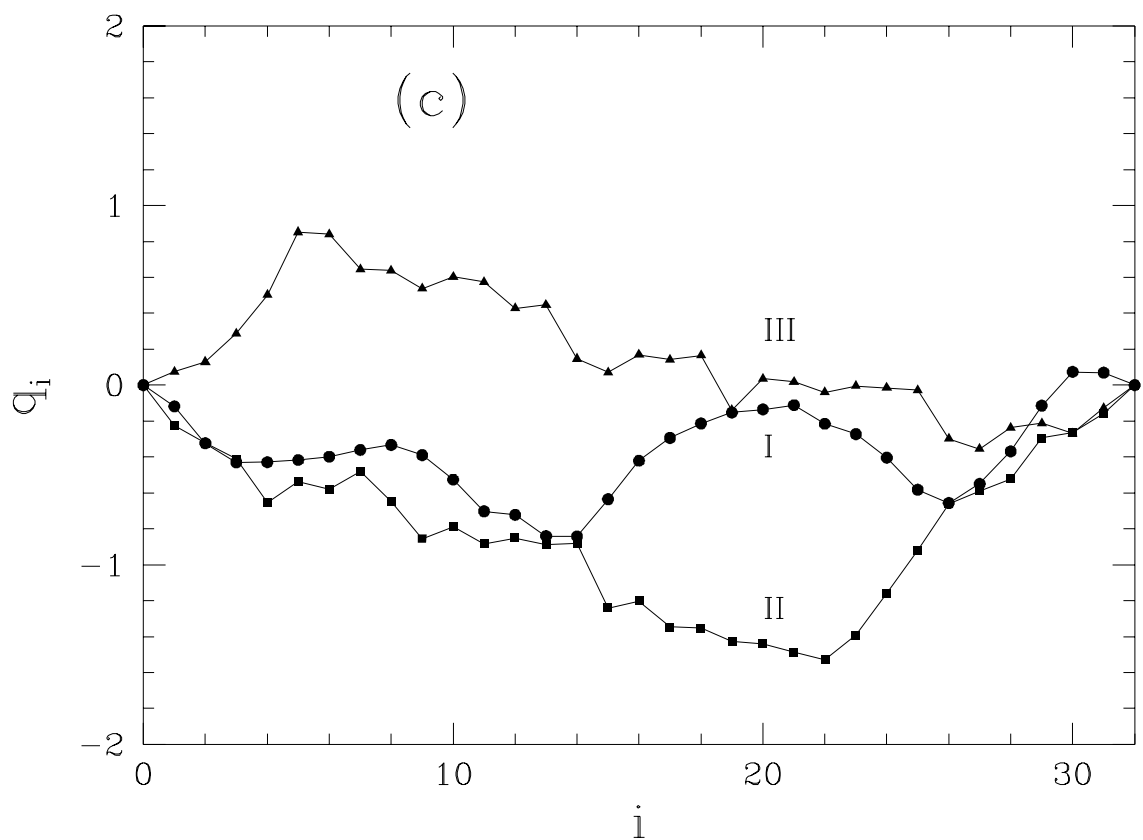
\begin{figure}
\caption{
Snapshots of the displacements $q_i$ of $32$ FPU-$\alpha$ particles as a 
function of position $i$ along the chain shown at different times. Here the 
initial amplitude is $A=3$ - as in Fig. 1 - for a sine wave [see Eq.(4)].
{\it (a)} ordered configurations at $t=0$ (I), $10^4$ (II) and $5\cdot 10^4$
(III); {\it (b)} ordered configurations at $t=10^5$ (I), $2\cdot 10^5$ (II) 
and $4.2\cdot 10^5$ (III), note 
the almost recurrence at $t=2\cdot 10^5$; {\it (c)} $t=1.6\cdot 10^6$ (I),
$5.4\cdot 10^7$ (II) and $2\cdot 10^8$ (III). 
Note the increasing disorder of the
configurations in {\it (c)}, corresponding to the onset of a chaotic dynamics
in the system, due to the breakup of regular soliton-like structures, and to 
its approach to equipartition. }
\end{figure}


\begin{references}
\bibitem{nota2} When the anharmonicity in the system is large, one should
define the weights on the basis of the generalized equipartition theorem,
for a discussion of this point see ref.{\protect\cite{PettiniCerruti}}.
\bibitem{pf}
The Poincar\'e-Fermi
non-existence theorem states that for a number $N$ of degrees of freedom
$\geq 3$, no analytic or smooth integral of motion besides the energy exists
for non-integrable systems. Thus, the non-integrable motions
(those that do not live on KAM tori) belong to a connected component of phase
space and the whole of any assigned constant energy surface of 
phase space is topologically accessible.
\bibitem{nota3}
We assume that even for the small amplitude considered by FPU a finite
threshold exists, so that we can extrapolate our results to the FPU case.
\end{references}
\end{document}